\documentclass[%
reprint,
 amsmath,amssymb,
 aps,nofootinbib
]{revtex4-2}
\usepackage{graphicx}
\usepackage{dcolumn}
\usepackage{bm}
\usepackage[colorlinks,
linkcolor=blue,       
anchorcolor=blue,  
urlcolor=blue,
citecolor=blue,        
]{hyperref}
\usepackage{placeins}
\usepackage{float}
\newcommand{\angstrom}{\textup{\AA}}

\begin{document}

\title{\large Spotting structural defects in crystals from the topology of vibrational modes}

\author{Long-Zhou Huang$^{1,2}$}
\author{Yun-Jiang Wang$^{1,2}$}%
 \email{yjwang@imech.ac.cn}
\author{Matteo Baggioli$^{3}$}
 \email{b.matteo@sjtu.edu.cn}
\affiliation{%
 $^{1}$State Key Laboratory of Nonlinear Mechanics, Institute of Mechanics, Chinese Academy of Sciences, Beijing 100190, China\\
 $^{2}$School of Engineering Science, University of Chinese Academy of Sciences, Beijing 101408, China \\
 $^{3}$Wilczek Quantum Center, School of Physics and Astronomy, Shanghai Jiao Tong University, Shanghai 200240, China 
}%


\begin{abstract}
Because of the inevitably disordered background, structural defects are not well-defined concepts in amorphous solids. In order to overcome this difficulty, it has been recently proposed that topological defects can be still identified in the pattern of vibrational modes, by looking at the corresponding eigenvector field at low frequency. Moreover, it has been verified that these defects strongly correlate with the location of soft spots in glasses, that are the regions more prone to plastic rearrangements. Here, we show that the topology of vibrational modes predicts the location of structural defects in crystals as well, including the cases of dislocations, disclinations and Eshelby inclusions. Our results suggest that in crystalline solids topological defects in the vibrational modes are directly connected to the well-established structural defects governing plastic deformations and present characteristics very similar to those observed in amorphous solids.
\end{abstract}   

\maketitle


\section*{Introduction}

Structural defects were postulated by Frenkel \cite{Frenkel1926} to explain theoretically why plastic yielding appears in crystals at a maximal stress that is thousand times smaller than what a naive estimate based on elasticity theory suggests. In three independent works dated 1934, Orowan \cite{orowan1934kristallplastizitat}, Polanyi \cite{Polanyi1934} and Taylor \cite{Taylor} identified the defects proposed by Frenkel with (edge) dislocation lines, that were already introduced by Weingarten \cite{weingarten1901surfaces} and Volterra \cite{volterra1907annls} a few decades before in the context of elasticity theory. 

\begin{figure}
    \centering
    \includegraphics[width=\linewidth]{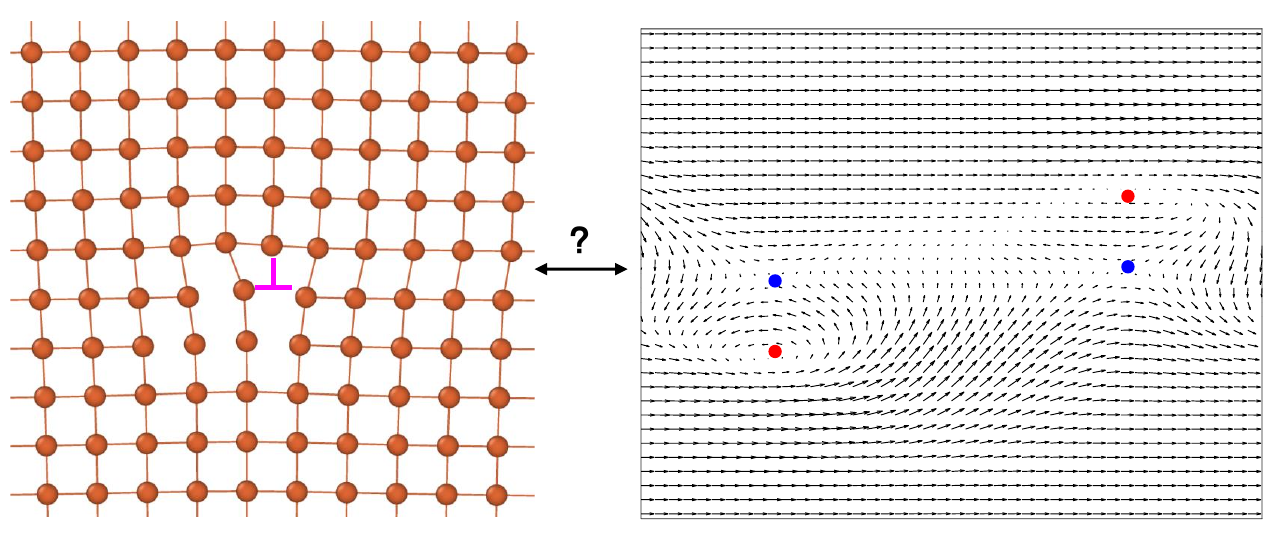}
    \caption{\textbf{Is there a correspondence between structural defects in crystals and topological defects in the vibrational modes?} The left image presents a 2D visualization of a dislocation line within an ordered atomic arrangement. The right figure depicts a typical structure for the eigenvector field of a low-frequency vibrational mode (eigenvector), where topological defects of positive and negative charge are indicated respectively with red and blue dots.}
    \label{fig:1}
\end{figure}

Dislocations now play a fundamental role in the description of mechanical deformation and plasticity in crystalline solids \cite{orowan1949fracture}. Despite dislocations, and also disclinations, can be formally defined using topology (homotopy classes) \cite{kleinert1989gauge} and geometrical concepts (torsion and Riemannian curvature respectively) \cite{nelson2002defects}, in ordered crystals these structural defects can be observed from direct imaging using for example scanning/transmission electron microscopy \cite{HYTCH1998131} or electron and x-ray diffraction-based techniques \cite{wheeler,cloete}, without many mathematical complications. In fact, there is no need of an expert eye to identify where the defect in the atomic structure presented in Fig.~\ref{fig:1} is located. Aside from direct imaging, a connection between vibrational modes and structural defects in crystals -- ``\textit{the sound of disorder}'' -- has also been discussed in the literature and verified experimentally \cite{RevModPhys.47.S1.2,PhysRevE.88.022315} (See also chapter VIII in \cite{maradudin1963theory} for a theoretical discussion of the effects of defects on lattice vibrations).

Disordered solids, such as glasses, do not display an ordered atomic arrangement as for crystals and, using technical jargon, do not possess long-range translational order \cite{binder2011glassy}. This complicates the definition of structural defects that is based on the existence of an ordered reference configuration \cite{baggioli2023topological}. Still, glasses, and in general amorphous solids, not only exhibit a strong plastic behavior but feature also the presence of ``soft spots'', regions where local atomic rearrangements are somehow more prone to happen. It is therefore common intuition that certain ``defects'' exist in glasses and that they are likely responsible for their plasticity and other anomalous properties \cite{doi:10.1142/q0371}. Unfortunately, until very recently the identification of these defective regions could only be performed using phenomenological structural indicators \cite{PhysRevMaterials.4.113609} or machine learning techniques \cite{PhysRevLett.114.108001} that, despite their success, do not provide the fundamental nature of these defects.

In the last years, important developments in the identification and characterization of topological defects in glasses have been made \cite{PhysRevLett.127.015501,wu2023topology}. The common idea behind these novel approaches is to look beyond structural properties and to search for topological fingerprints in other ``dynamical'' physical quantities. More in detail, Wu et al. \cite{wu2023topology} suggested that the topology of the vibrational modes, encoded in the spatial pattern of the low-frequency eigenvector field, reveals vortex-like topological defects that accurately predict the location of plastic events in glasses. Some of these vortex-like defects have been related to the displacement fields expected of an Eshelby inclusion \cite{PhysRevE.109.L053002}. Vortex-like structures have also been recently connected to the formation and dynamics of shear bands in glasses \cite{Sopu2017,PhysRevB.110.014107}. Importantly, these topological defects in the vibrational modes have been experimentally observed in a 2D colloidal glass \cite{vaibhav2024experimental}. Moreover, they have been also shown to play an important role in 3D glasses \cite{10.1093/pnasnexus/pgae315}, where their definition has to be extended to so-called Hedgehog defects in order to retain their topological character \cite{bera2024hedgehog}.

In summary, numerous results have now confirmed the promising role of topological defects in the vibrational modes and their strong correlation with plasticity in glasses. Nevertheless, the physical significance of these defects and their, if any, connection with structural properties remain largely unexplored.

In this work, we take a step back and ascertain whether these newly discovered topological objects bear any connection with well-established structural defects in crystals, such as dislocations, disclinations and Eshelby inclusions (see Fig.~\ref{fig:1} for a visual representation of our main question). As already noted, there is no need to resort to any of these new methods to locate structural defects in ordered crystals. Nevertheless, comparing them with well-understood concepts will help illuminating their physical significance and will provide a firmer ground to the recent claims that these dynamical defects generalize the concept of structural defects to disordered structures.  

Spoiling the conclusions of our manuscript, we anticipate that the newly defined topological defects, identified in the lowest frequency eigenvector field, nicely spot structural defects in 2D crystalline solids. Moreover the structural and statistical characteristics of these topological defects appear surprisingly similar in amorphous solids and defective crystals. The rest of this manuscript will be devoted to prove these claims.

\section*{Topology of vibrational modes}

In solid systems, the study of vibrational dynamics relies on the normal mode analysis \cite{moon2023normal} that is based on the computation of the Hessian matrix,
\begin{equation}
    H_{ij} = \frac{1}{\sqrt{m_im_j}} \frac{\partial^2U}{\partial r_i \partial r_j},
\end{equation}
where $m_i$ and $r_i$ are the mass and position of the $i$th particle and $U$ the total potential energy of the system. By direct diagonalization of the Hessian matrix, one obtains the eigenvalues $\lambda_j$ and the corresponding eigenvectors $\vec{e}_j$, where $j$ is a discrete index over the various vibrational modes. The eigenfrequencies are then derived from the relation $\lambda_j\equiv \omega_j^2$. In this way, each mode is associated to a frequency and to a corresponding eigenvector. In the rest of the manuscript, we will consider two-dimensional (2D) systems for which the eigenvector fields $\vec{e}_j$ are 2D vectors spanning over coordinates $x$, $y$.

\begin{figure*}
    \centering
\includegraphics[width=\linewidth]{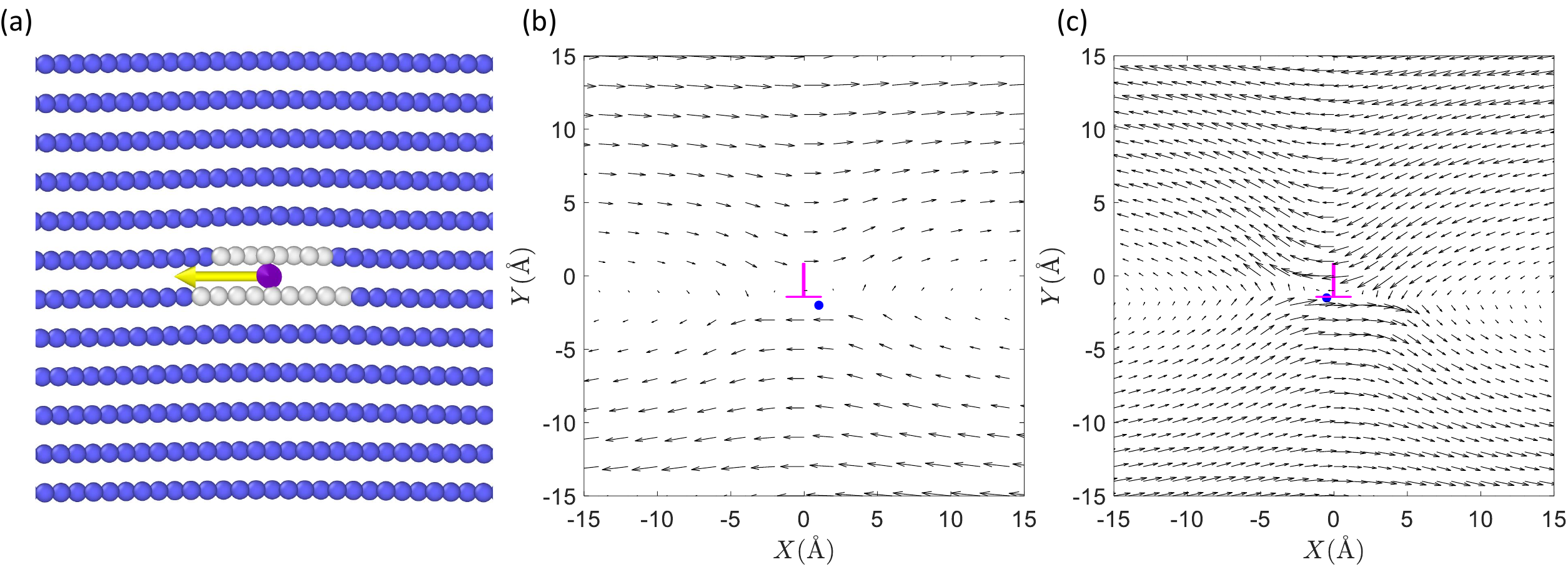}
    \caption{\textbf{A single dislocation.} \textbf{(a)} A full edge dislocation in a BCC iron crystal. Blue atoms represent the BCC crystalline lattice while white atoms locate the edge dislocation core. The dislocation line perpendicular to the paper plane is represented in purple color. The yellow arrow represents the Burgers vector and thus the slip direction of the dislocation. \textbf{(b)} The smoothed-out eigenvector field for the lowest mode with frequency $\omega=0.63$ THz. In this panel, a relatively coarse grid size of $\mathrm{d} r = 2$ $\angstrom$ is used to compute the topological charge. \textbf{(c)} The smoothed-out eigenvector field (with $\mathrm{d} r = 1$ $\angstrom$) for the second lowest mode with frequency $\omega=0.97$ THz. The blue symbol locates the position of anti-vortex. The magenta $\perp$ symbol indicates the edge dislocation.}
    \label{fig:2}
\end{figure*}

In 2D, we then define the topological charge (winding number),
\begin{equation}
    q=\frac{1}{2\pi}\oint_{\mathcal{C}} d\theta,\label{wind}
\end{equation}
where $\mathcal{C}$ is a closed curve of arbitrary shape and $\theta$ a continuous angle that determines the direction of the eigenvector field at each point in the 2D plane. The topological charge $q$ is quantized and takes integer values $\pm n$. In particular, $q=+1$ corresponds to a topological vortex, while $q=-1$ to a topological anti-vortex. We notice that $q$ is a topological charge and is therefore insensitive to continuous (non-singular) transformations, \textit{i.e.}, the same reason why a mug and a donut are topologically equivalent.

The eigenvector field obtained from simulations is defined only at the discrete particle positions (see \cite{PhysRevE.103.012602} for a discussion about the analysis of topological defects in discrete vector fields). Therefore, an interpolation procedure is necessary to arrive at a continuous angle $\theta$. Following \cite{wu2023topology}, we define the phase angle $\theta (\vec{r})$ at each lattice site as
\begin{equation}
    \tan\theta(\vec{r})=\frac{\sum_i w(\vec{r}-\vec{r}_i)e^y_i}{\sum_i w(\vec{r}-\vec{r}_i)e^x_i},
\end{equation}
where $\vec{r}_i$ is the location of particle $i$, and $w(\vec{r}-\vec{r}_i)$ is a Gaussian weight function,
\begin{equation}
    w(\vec{r}-\vec{r}_i)=\exp(-|\vec{r}-\vec{r}_i|^2/r_c^2)
\end{equation}
with $r_c=1\ \angstrom $. We then evaluate the winding number $q$, Eq.~\eqref{wind}, using as closed circuit $\mathcal{C}$ the smallest square path on the lattice. Each square giving a $\pm 1$ winding number is finally associated to a vortex (anti-vortex) located at its center. Using this systematic method, we can locate the topological defects in the eigenvector field of each eigenmode.

In the rest of the manuscript, we will make use of several atomistic simulation models. For the details on the models and the simulations, we refer the Reader to the Methods sections \ref{methodsa} and \ref{methodsb}. It should be noted that in the rest of this manuscript, we will always use a grid size of $1\ \angstrom$. Panels (b) in Fig.~\ref{fig:2} is an exception, where the grid size is taken to be $2\ \angstrom$ to demonstrate a coarse-grained effect in the analysis of the topological charge. The robustness of our results upon dialing the cutoff distance and the grid size is confirmed in Appendix \ref{add}.

\section*{Dislocations}

Dislocations are topological defects of translational nature \cite{hull2011introduction} that can be formally defined using a topological quantity known as the Burgers vector \cite{anderson2017theory},
\begin{equation}
    \vec{b}\equiv - \oint_\mathcal{L} d\vec{u},
\end{equation}
where $\vec{u}$ is the displacement field and $\mathcal{L}$ a closed circuit. A finite Burgers vector, signaling the presence of a dislocation, is equivalent to a violation of the compatibility constraint for the strain tensor \cite{love2013treatise}. Dislocations can be classified into screw and edge dislocations depending whether the dislocation line and the Burgers vector are parallel or perpendicular to each other. 

\textit{Can topological defects in the vibrational modes locate the position of a dislocation in a 2D crystal}?\footnote{See \cite{jiang2024singleatomresolvedvibrationalspectroscopydislocation} for a recent experimental study using electron energy-loss spectroscopy on a GaN dislocation.} In order to answer this question, we consider the example of an edge dislocation in a quasi-two-dimensional BCC iron crystal. In panel (a) of Fig.~\ref{fig:2}, we show a snapshot of the simulation system of a single dislocation in iron where blue atoms represent the ordered BCC atomic arrangement, while the white atoms locate the dislocation core. The yellow arrow represents the Burgers vector and thus the edge dislocation motion direction. The purple dot represents the position of the dislocation line.

Panels (b) and (c) in Fig.~\ref{fig:2} show the spatial pattern of the 2D eigenvector field corresponding to the lowest vibrational modes with frequency $\omega= 0.63$ THz and $\omega=0.97$ THz, respectively. Away from the dislocation core, we observe that both modes display a smooth and ordered structure with no evident singularities. On the contrary, close to the dislocation core (indicated with a magenta $\perp$) the eigenvector field shows a quite intricate pattern. The blue dot indicates the location of an anti-vortex topological defect with negative charge $q=-1$. As evident from panels (b) and (c) in Fig.~\ref{fig:2}, the location of the anti-vortices in the two lowest frequencies vibrational modes coincides up to numerical precision with the core of the edge dislocation. 

\begin{figure}
    \centering
    \includegraphics[width=1\linewidth]{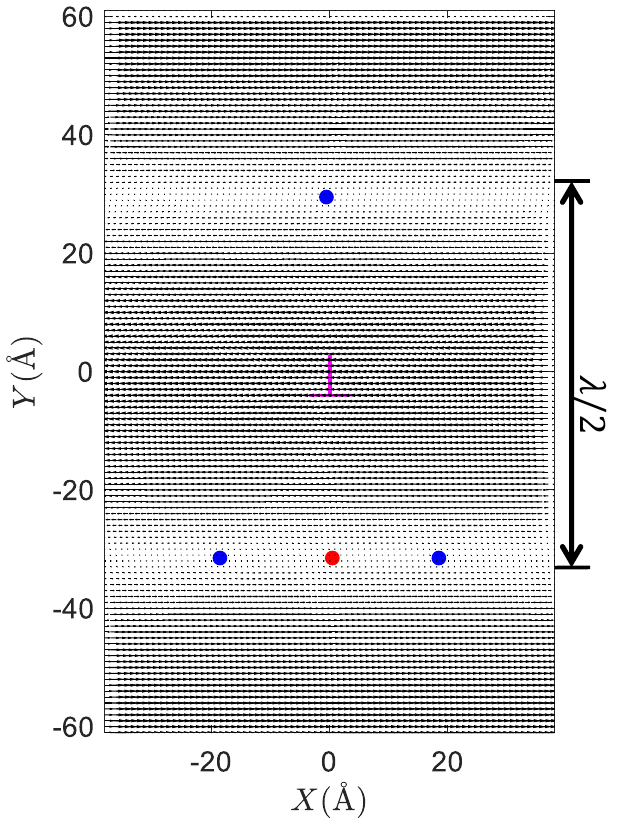}
    \caption{\textbf{Not all topological defects are equal.} The spatial structure of the third lowest vibrational modes ($\omega=1.28$ THz) for the case of a single dislocation depicted in Fig.~\ref{fig:2}. The dislocation core is indicated with the $\perp$ symbol. Positive and negative topological defects are labelled respectively by red and blue dots. The black arrow on the right indicates the characteristic length scale $\lambda/2$ where $\lambda$ is the wavelength of the transverse phonon at this frequency.}
    \label{fig:a1}
\end{figure}

Higher frequency vibrational modes present a more complex pattern where topological defects proliferate. This proliferation and the structural properties of the topological defects as a function of $\omega$ will be discussed in more detail in the next sections. Nevertheless, it is important to already stress that many of the defects appearing in the higher frequency vibrational modes are not directly related to the structural defects, in this case the single edge dislocation. This is reasonable since most of the high-frequency modes are not relevant to the plasticity of crystals; in fact, it is usually the lowest mode that governs the dislocation motion under applying external stress \cite{Wang2023}. The proliferation of topological defects with increasing frequency is evident from the structure of the third lowest vibrational mode shown in Fig.~\ref{fig:a1}. There, we can observe the presence of four distinct topological defects far away from the dislocation core. These defects are located along two bands at approximately $Y=\pm 30$ $\angstrom$. From the ordered vibrational pattern, it is evident that these two bands are of phononic nature. Indeed, it is well-known that the superposition of transverse phononic modes can give rise to an ordered pattern of topological defects in the vibrational modes (see for example Fig. 3 in \cite{lerner2024generic} or Fig. 3 in \cite{doi:10.1142/9781800612587_0002}). To confirm that, we have computed the transverse speed of sound $c_T$ and the corresponding wavelength $\lambda=2 \pi c_T/\omega$. For the third mode in Fig.~\ref{fig:a1}, we find that $\lambda \approx 126.85$ $\angstrom$. As indicated by the arrow on the right, the distance between the two bands, where the topological defects appear, coincides exactly with $\lambda/2$, confirming their origin from transverse phononic vibrations. 

\begin{figure}
    \centering
\includegraphics[width=1\linewidth]{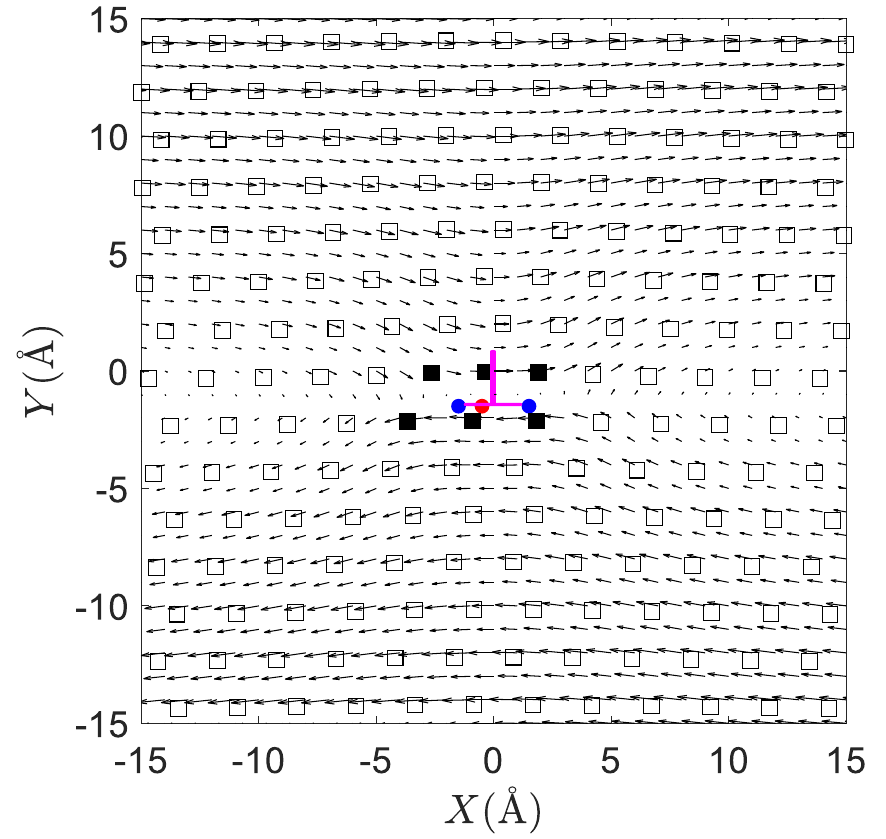}
    \caption{\textbf{Resolving the internal topological structure of a single dislocation.} A zoom of the smoothed-out eigenvector field for the lowest mode, $\omega=0.63$ THz, corresponding to the dislocation in Fig.~\ref{fig:2}(a). Here, compared to Fig.~\ref{fig:2}(b), a much smaller grid is used to resolve the internal structure of the dislocation core (black solid squares, $\blacksquare$), whose size is about $4.5$ $\angstrom$. Blue and red symbols are respectively anti-vortices and vortices.}
    \label{fig:4}
\end{figure}

Furthermore, it is important to stress that Eq.~\eqref{wind} computes the total topological charge inside the loop $\mathcal{C}$ that, in general, might correspond to a combination of many microscopic topological defects located inside $\mathcal{C}$. For example, a closed loop surrounding a vortex-antivortex pair would result in $q=0$. This implies that, when relatively large grid sizes are used, the definition of microscopic defects becomes subtle and one should rather discuss local topological charge. Given these facts, it is therefore instructive to use the smallest grid available to resolve as much as possible the microscopic topological structure of the vibrational modes. In Fig.~\ref{fig:4}, we show the results of this analysis for the lowest mode with $\omega=0.63$ THz corresponding to the single dislocation depicted in Fig.~\ref{fig:2}(a). We used a grid whose size ($1$ $\angstrom$) is much shorter than that of the dislocation core (about $4.5$ $\angstrom$), allowing us to explore the interior structure of the dislocation and its topological characteristics. The atoms that are part of the dislocation core are colored black solid ($\blacksquare$) for better visualization; the other atoms are indicated with empty squares ($\square$). By performing this analysis with the smallest grid, we identify inside the dislocation core a dipole of negative and positive defects, that are very close to each other, and an additional negative topological defect nearby. Consistent with our previous findings, the total topological charge computed in a loop $\mathcal{C}$ surrounding the entire dislocation core is still $q=-1$, as previously revealed using a larger grid size in Fig.~\ref{fig:2}(b). In summary, the total topological charge associated with a single dislocation is always $q=-1$, but the internal structure within the dislocation core is richer than a single negative topological defect. More details about the dependence of $q$ on the grid size have been further demonstrated in the Fig.~\ref{fig:10} in Appendix \ref{add}.

\begin{figure*}
    \centering
    \includegraphics[width=\linewidth]{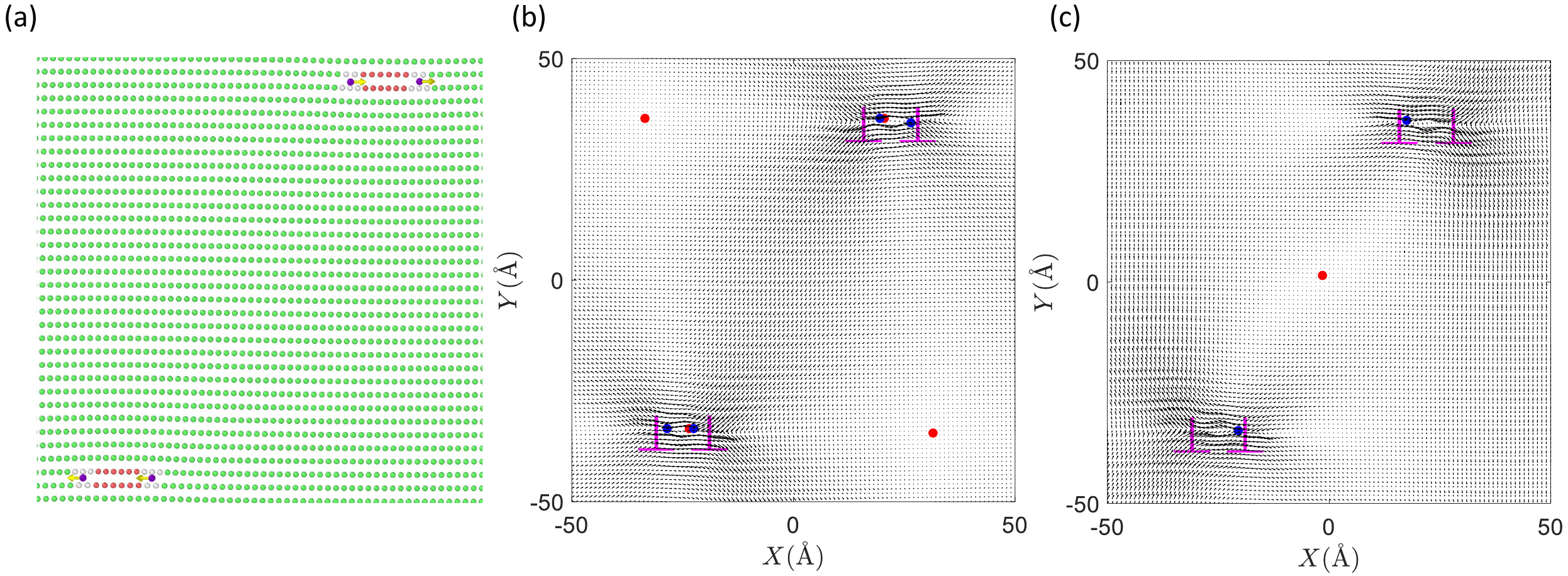}
    \caption{\textbf{Partial dislocations and stacking faults.} \textbf{(a)} A FCC aluminum crystal (green atoms) with four partial dislocations (white atoms) and stacking faults (red atoms). The yellow arrows denote the Burgers vectors and thus the slip direction of edge dislocations. \textbf{(b)} The smoothed-out eigenvector field for the lowest mode with frequency $\omega=0.32$ THz. \textbf{(c)} The eigenvector of the second lowest mode, $\omega=0.84$ THz. The red and blue symbols locate vortices and anti-vortices, respectively. The magenta $\perp$ symbols indicate the four partial dislocations.}
    \label{fig:5}
\end{figure*}

In order to confirm the validity of our results, in Fig.~\ref{fig:5} we consider a more complex situation where four partial dislocations are embedded in an aluminum crystal of FCC. The four partial $1/6\,\langle1\ 1\ 2\rangle$ dislocations are decomposed from two full $1/2\,\langle1\ 1\ 1\rangle$ dislocations, with a stacking fault in between connecting the partials. In panel (a) of Fig.~\ref{fig:5}, we show an image of the crystal where green atoms represent the FCC lattice, white atoms the partial dislocation cores, and red atoms the stacking faults connecting them. The consideration of a full dislocation dipole (four partials) in aluminum facilitates using periodic boundary conditions in the simulation box. 

In panel (b) of Fig.~\ref{fig:5} we show the spatial pattern of the eigenvector of the lowest frequency mode with $\omega=0.32$ THz. We observe two distinct features. First, each of the four partial dislocations (white atoms in Fig.~\ref{fig:5}(a)) is associated with a topological charge $-1$, indicated by blue dots. This is consistent with our previous findings related to a single dislocation, Fig.~\ref{fig:2}. In addition to that, we notice the presence of four $q=+1$ vortices aligned in a regular pattern. Two of them appear far away from the structural defects in panel (a), suggesting that their origin might not be directly related to the partial dislocations or the stacking faults connecting them. Unfortunately, in this case, we are not able to determine with accuracy the transverse speed of sound by shearing the system due to the strong interactions between the partial dislocations. Moreover, given the high density of defects, we do expect a strong interaction between extended plane wave vibrations and the dynamics related to the defects. We nevertheless attempt a crude estimate of the phonon wavelength, that gives $\lambda\approx 248$ $\angstrom$. The diagonal distance between the orderly arranged vortices is approximately $113$ $\angstrom$, and it is not so far from $\lambda/2$. We also notice that the two vortices appearing next to the partial dislocations are slightly misaligned. This is probably a result of the interaction with the structural defects. In summary, also in this case, we have evidenced that the additional vortices in the lowest mode arise because of the interactions between phononic vibrations and the structural defects. In particular, the ordered square arrangement of the four vortices and their distance very close to $\lambda/2$ support this interpretation.

\begin{figure*}
    \centering
    \includegraphics[width=0.9\linewidth]{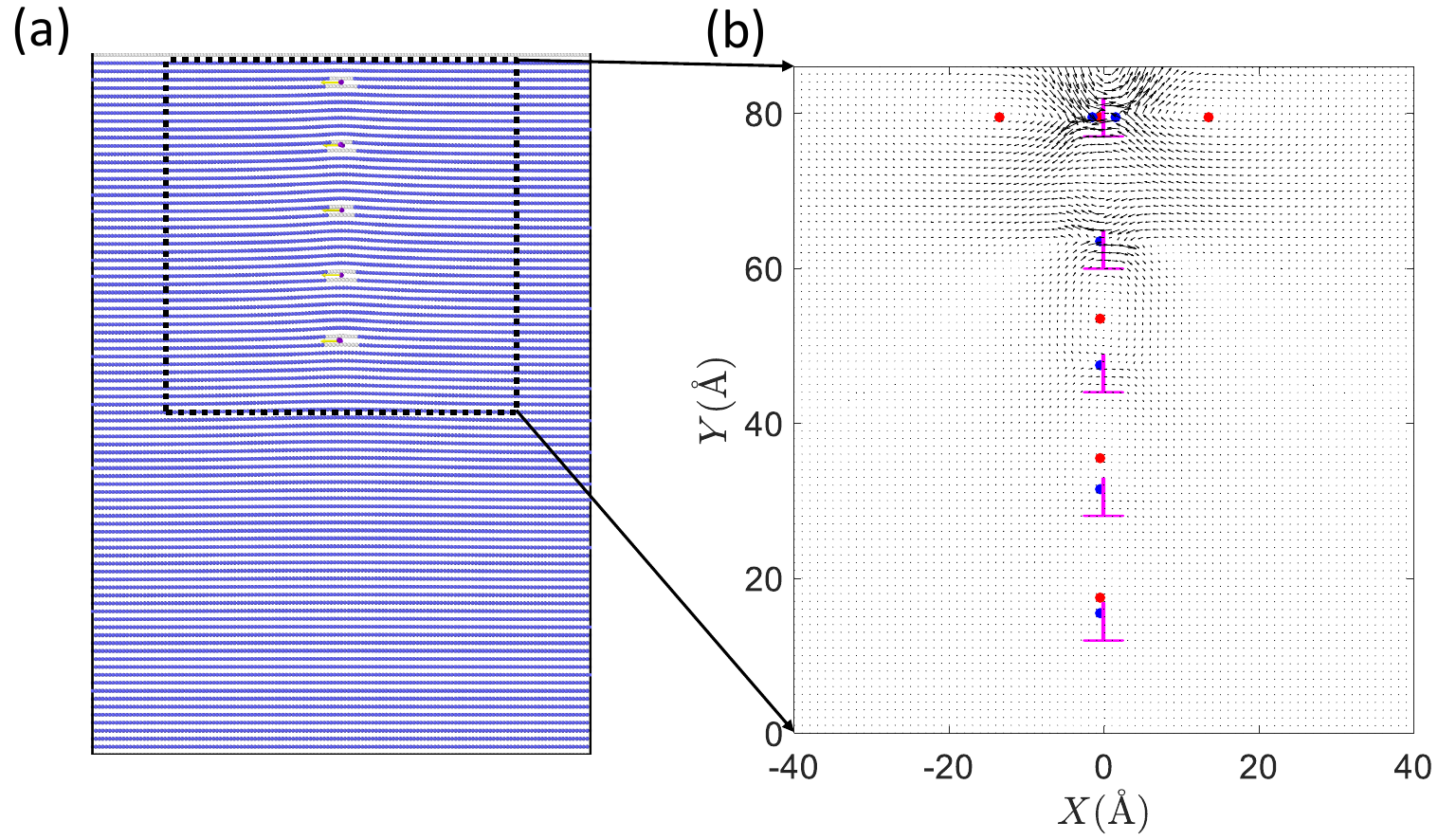}
    \caption{\textbf{A single disclination.} \textbf{(a)} Top view of an array of 1/2 $\langle$1\ 1\ 1$\rangle$ edge dislocations in a iron crystal that forms a wedge disclination. \textbf{(b)} The eigenvector of the lowest frequency vibrational mode with $\omega=0.43$ THz. Red and blue symbols are respectively vortices and anti-vortices. }
    \label{fig:6}
\end{figure*}

In panel (c) of Fig.~\ref{fig:5}, we show the structure of the second lowest mode, $\omega=0.84$ THz. In here, we observe that the identification of the structural defects using the topology of the vibrational mode becomes less transparent. In particular, we notice that the region near the partial dislocation is highly distorted, but we can only identify a negative topological charge, indicated by the respective blue dot. Interestingly, we also observe a positive defect in the middle of the line connecting the two partial dislocation pairs. We expect that this defect is not directly related to the structural defects. It is either of phononic nature, as the vortices in panel (b) of Fig.~\ref{fig:5}, or just a result of the frustration between the two disordered regions in the top right and bottom left corners.

\section*{Disclination}

After studying in detail the case of dislocations in the previous section, we now move to the case of disclinations. It is well-known that dislocations and disclinations are not independent in crystals \cite{friedel1964dislocations}. Disclinations are topological defects of rotational nature \cite{RevModPhys.80.61} that are defined using a topological quantity known as the Frank vector,\footnote{In 2D, similarly to the magnetic field, the Frank vector becomes a pseudoscalar.}
\begin{equation}
    \vec{\Omega}=\oint_\mathcal{L} d\vec{\upsilon} \quad \text{with} \quad \vec{\upsilon}=\vec{\nabla} \times \vec{u}.
\end{equation}

In Fig.~\ref{fig:6}(a), we show a single disclination in a BCC iron crystal. We notice that the disclination is equivalent to an array of edge dislocations, whose cores are indicated in white color. In this sense, this disclination is equivalent to a grain boundary with a small tilt angle. In panel (b) of Fig.~\ref{fig:6}, we show the lowest vibrational mode with $\omega=0.43$ THz. Once again, we observe that each dislocation is accompanied by the presence of a negative topological charge, indicated by the blue dots. Interestingly, we also observe the presence of additional $+1$ vortices in between the dislocations and close to the top boundary of the system, which is a free surface of the crystal. These $+1$ defects are not related to phononic vibrations since the wavelength of the phonon modes with $\omega=0.43$ THz is given by $\lambda \approx 379$ $\angstrom$ and is much larger than the distance between the vortices. We interpret the appearance of these additional structure as follows. As shown in Fig.~\ref{fig:2}, a single and isolated dislocation is characterized by a $-1$ topological charge. Nevertheless, when many dislocations are arranged close to each other, they strongly distort the vibrational pattern in between them, inducing the appearance of more topological defects as a result of this frustration. We also notice that near the surface the structure of the lowest mode becomes more complex, as it is probably affected by finite size boundary effects. The topological defect, phonon, and surface modes hybridize each other resulting in a complex scenario of vibrational features as shown in Fig.~\ref{fig:6}(b).   

We can conclude that the case of a single dislocation (Fig.~\ref{fig:2}) presents a clearer connection to the topological structure of the lowest vibrational mode. A single disclination is more similar to the case of the partial dislocations, where the vibrational pattern becomes more complex and the connection to topology less direct. In general, we can affirm that scenarios where many dislocations coexist close to each other are harder to interpret only looking at the position of the topological defects in the lowest vibrational modes. This is consistent with the assumption that is the interaction between neighboring defects that cooperatively drives the onset of plasticity in these cases. 

\section*{Eshelby inclusion}

Here, we consider another type of defect that has been widely discussed in the literature, both in the context of crystalline and amorphous solids. More precisely, we discuss the case of Eshelby inclusion \cite{Eshelby1959}, that is often adopted as a toy model for quadrupolar localised defects. We clarify that, differently from dislocations and disclinations, Eshelby inclusions are not topological defects since no topological charge can be directly associated to them. For completeness, we consider two cases, \textit{i.e.}, a soft and a hard Eshelby inclusion in terms of iron inclusion in carbon matrix, and vice versa.

\begin{figure*}
    \centering
    \includegraphics[height=\textheight]{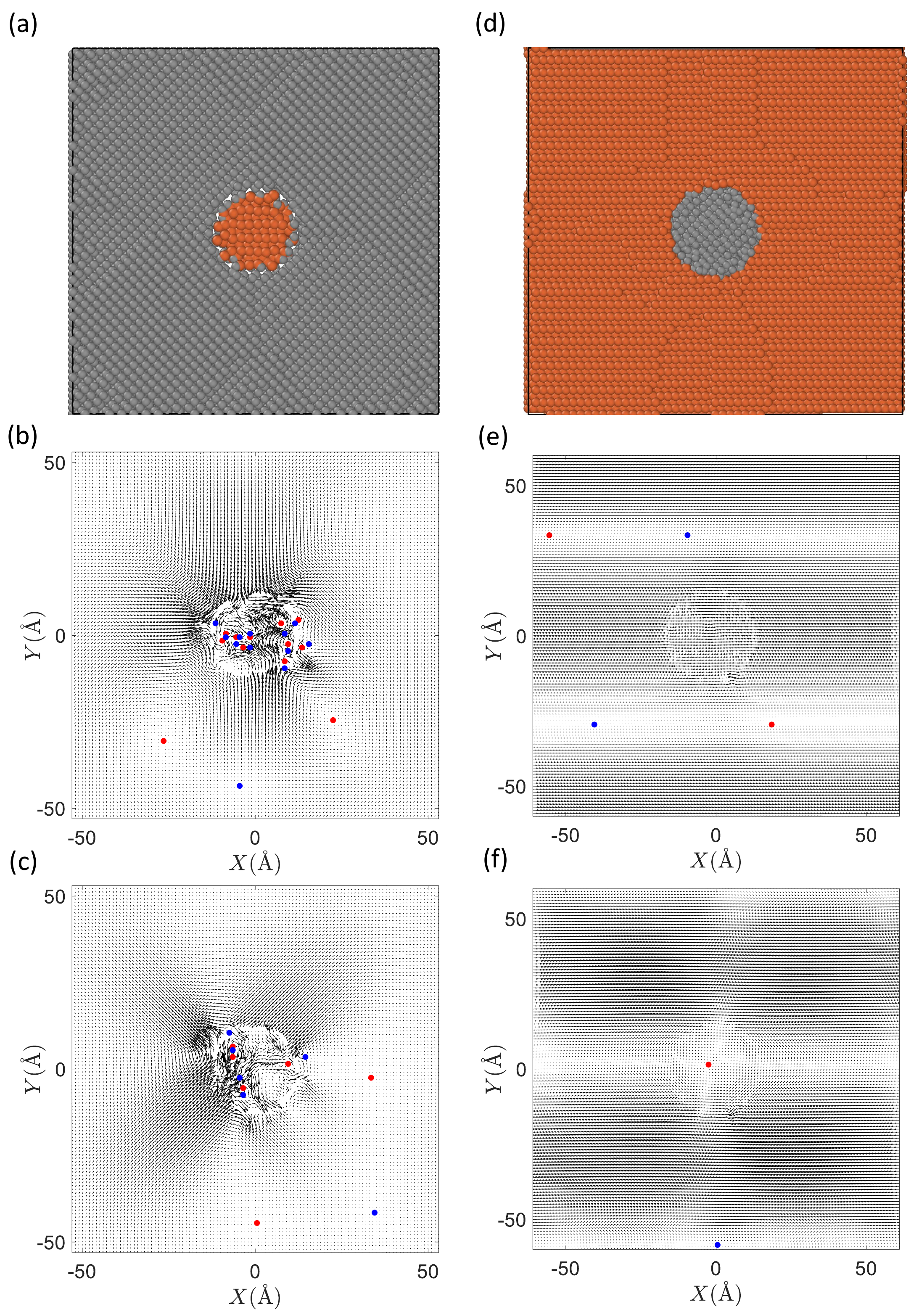}
    \caption{\textbf{Soft and hard Eshelby inclusions.} \textbf{(a)} Top view of a soft BCC iron inclusion in carbon of diamond lattice. \textbf{(b)-(c)} Eigenvector of the two lowest frequency modes with $\omega=2.76, 3.18$ THz and the corresponding topological defects. \textbf{(d)-(f)} The same plots for a hard carbon inclusion in a BCC iron. The two soft modes are with frequencies $\omega=1.11, 1.15$ THz. Blue and red symbols are respectively the anti-vortices and vortices. We notice that the total topological charge is always zero.}
    \label{fig:7}
\end{figure*}

First, we demonstrate the scenario of vibrational defects in a soft Eshelby inclusion, where a soft circular iron inclusion is embedded in the hard carbon matrix, as shown in the model in Fig.~\ref{fig:7}(a). The gray atoms reflect the carbon lattice while the brown ones correspond to the iron cylinder. Since the modulus of iron is much smaller than that of carbon, it is expected that the soft inclusion should present features of quadrupolar symmetry in the stress or strain field \cite{Eshelby1959,Maloney2006,Lemaitre2014,Hieronymus-Schmidt2017}. Here, this expectation is realized in Fig.~\ref{fig:7}(b) and (c) where the eigenvector structures for the two softest modes ($\omega=2.76, 3.18$ THz) do indeed exhibit quadrupolar patterns. In panel (b), the shear directions are horizontal and vertical for the lowest frequency mode; while shear is along diagonal directions for the second lowest frequency mode as shown in panel (c). 

We notice that the interior of the soft Eshelby inclusion presents a complex pattern of topological defects. Interestingly, we find that (I) the total topological charge computed surrounding the whole inclusion is $q=-1$ for both the first and second lowest frequency modes, and (II) most of the defects inside the inclusion are arranged in pairs, forming dipoles of $+1$ and $-1$ charges.
Importantly, our finding, associating a $-1$ topological charge to the quadrupolar inclusion, is in agreement with the results recently reported in shear transformation of glasses \cite{PhysRevE.109.L053002}, although in \cite{PhysRevE.109.L053002} the topological defects were obtained from the dynamical displacement field while ours from the vibrational modes. It is important to notice that the dynamical displacement field is built on an infinite sum over all the eigenmodes, however, the most important contribution is always from the soft modes. Therefore it is not surprising that the quadrupolar symmetry from vibration coincides with that from displacement field, although in the present case of inclusion in crystals, the mechanics is highly anisotropic, whereas it is isotropic in the case of glasses. It is nevertheless interesting to observe that both the topology of the displacement field and that of the lowest vibrational modes can locate the presence of an Eshelby inclusion.


In the case of hard Eshelby inclusion as shown in Fig.~\ref{fig:7}(d)-(f), the topological defects are trivial. The phononic vibration of the soft iron matrix is not affected by the existence of the hard carbon inclusion. It is indeed evident that the topological defects in Fig.~\ref{fig:7}(e)-(f) are not caused by the inclusion but they are of phononic nature.

Finally, we observe that for both soft and hard inclusions (and differently with the case of dislocations and disclinations) the total topological charge in the whole system is always zero. In fact, there are always pairs of negative and positive defects as seen from eigenvector of the softest modes in panels (b), (c), (e) and (f) of Fig.~\ref{fig:7}. This charge neutrality condition is a direct consequence of the fact that Eshelby inclusions are not topological, as stated at the beginning of our discussion.  The findings also apply to higher frequency modes as demonstrated in Fig.~\ref{fig:11} of Appendix \ref{add}.

\begin{figure}
    \centering
    \includegraphics[width=\linewidth]{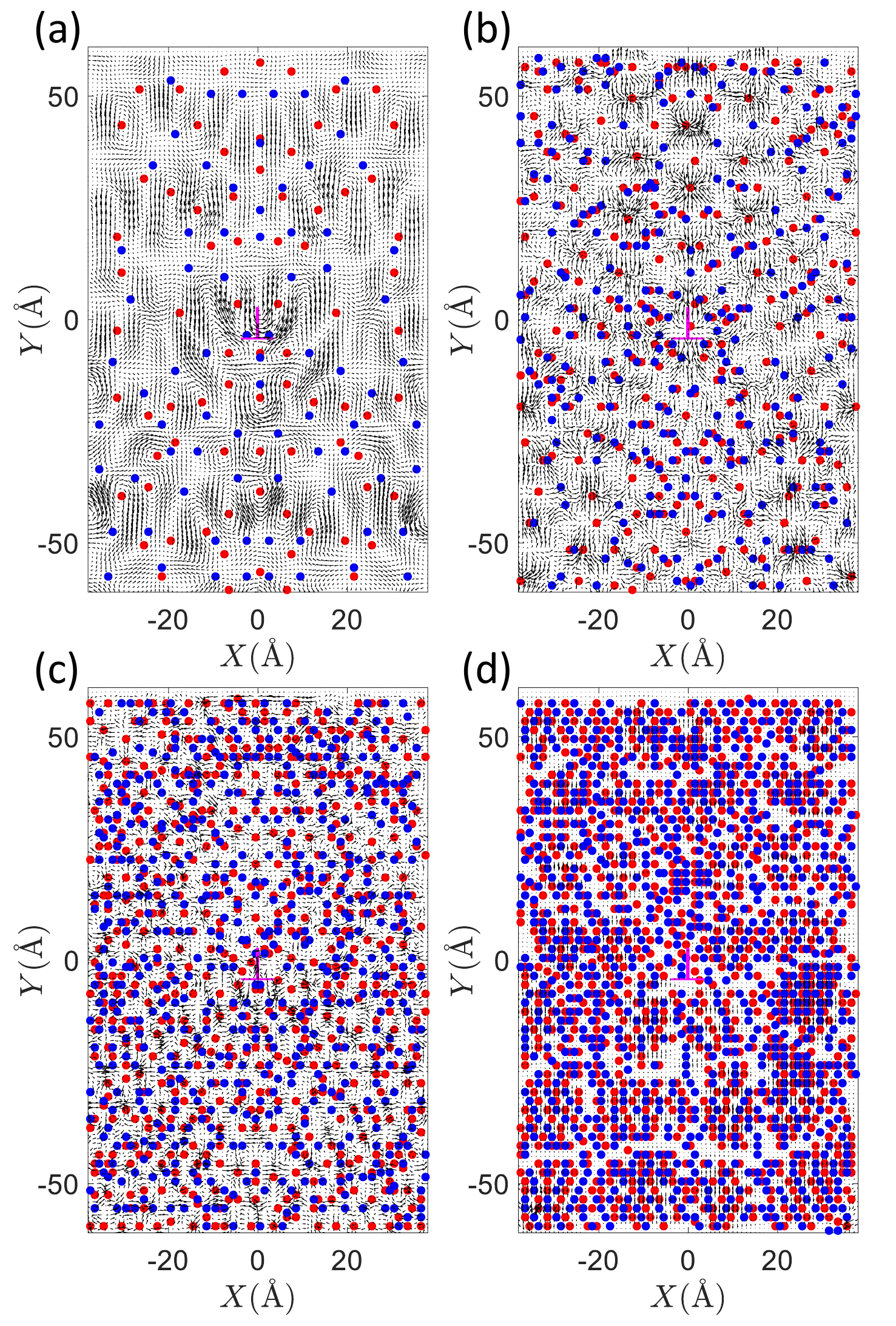}
    \caption{\textbf{Proliferation of topological defects in the higher frequency vibrational modes.} The eigevenctor field for the higher frequency vibrational modes $\omega=15,\,30,\,45,\,60$ THz and the location of the topological defects for the case of a single dislocation in Fig.~\ref{fig:2}.}
    \label{fig:8}
\end{figure}

\begin{figure*}
    \centering
    \includegraphics[width=\linewidth]{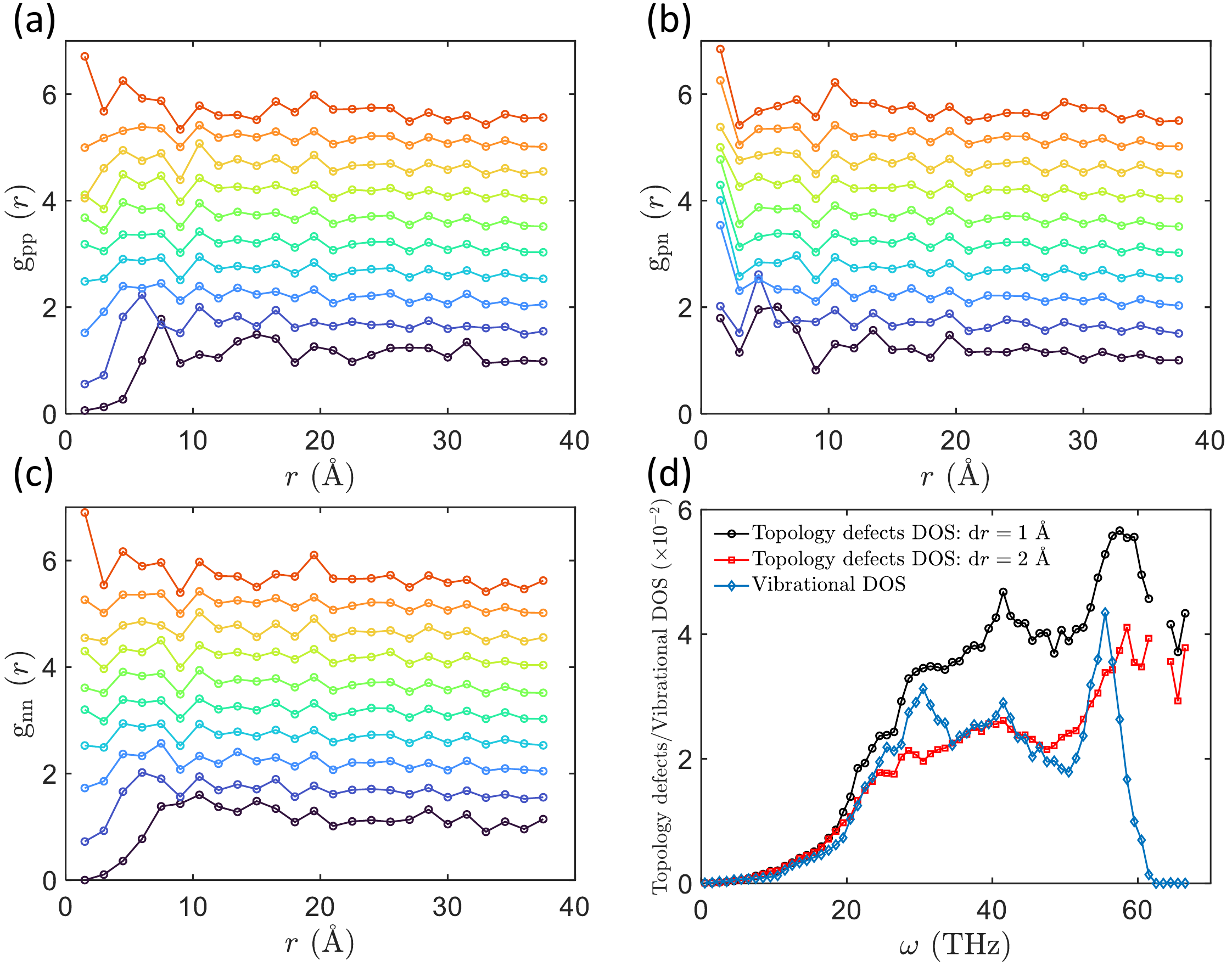}
    \caption{\textbf{Characteristics of topological defects.} Panels \textbf{(a)-(c)} Pair correlation functions for positive-positive, positive-negative and negative-negative defect pairs ($g_{pp}(r), g_{pn}(r), g_{nn}(r)$), respectively, for different frequencies $\omega=15, 20, 25, 30, 35, 40, 45, 50, 55, 60$ THz from bottom to top. Plots for $\omega>15\ \mathrm{THz}$ are shifted vertically by the multiple of 0.5 for clarity.
 \textbf{(d)} Number of topological defects as a function of the frequency $\omega$ at different sizes of grid $\mathrm{d}r$ for the single edge dislocation in Fig.~\ref{fig:2} and corresponding density of states (DOS).}
    \label{fig:9}
\end{figure*}

\section*{Topological defects in higher frequency vibrational modes}

So far, we have focused our attention of the topology of the several lowest vibrational modes since we do expect them to be the most relevant to predict and describe the plastic deformation corresponding to the structural defects of interest. Nevertheless, it is interesting to explore how the topology of the vibrational modes evolve by increasing the eigenfrequency $\omega$.

In Fig.~\ref{fig:8}, we show the structure of the higher-frequency eigenmodes for the case of a single dislocation presented in Fig.~\ref{fig:2}. It is immediately evident that the number of topological defects increases rapidly with the frequency. Moreover, the structure of the eigevenctors becomes more random, the position of the defects decorrelate and the swirling structures disappear. This tendency is qualitatively identical to what found in amorphous solids in \cite{wu2023topology} (see also \cite{vaibhav2024experimental,bera2024hedgehog}). Moreover the positional correlation of the topological defects is gradually lost and their structure at high frequency resembles more that of a decorrelated gas of defects, as evident from panel (d) in Fig.~\ref{fig:8}.

In order to better quantify the proliferation of the topological defects by increasing the frequency, we provide a further analysis on their spatial correlations and number density.

First, we define the partial radial distribution functions between the topological defects in each eigenvector field,
\begin{equation}
    g_{\alpha\beta}^\omega(r)= \frac{S}{2 \pi r \Delta rN_{{\alpha }} N_{{\beta }}} \sum_{i=1}^{N_{{\alpha }}} \sum_{j=1}^{N_{{\beta}}} \delta\left(r-\left|\vec{r}_{i j}\right|\right), 
\end{equation}
where $\alpha,\beta=\{\mathrm{p},\mathrm{n}\}$ (p = positive, n = negative) and $\omega$ is the eigenfrequency of the corresponding eigenvector, $S$ is the area of the 2D region, $N_{{\alpha,\beta }}$ are the number of topological defects of the eigenfrequency $\omega$ . At low frequency, \textit{e.g.}, $\omega=15$ THz in Fig.~\ref{fig:9}, we observe no correlation (or a correlation hole) between positive-positive and negative-negative topological defects at short distance, $r\lesssim 5$ $\angstrom$. This implies that at short distance defects with same sign of their charges tend to repel each other. On the contrary, in the same low-frequency regime, we observe a strong correlation between positive and negative defects suggesting a tendency to pairing and clustering. These features are analogous of what found in glasses in \cite{wu2023topology,vaibhav2024experimental}. At low frequency, the correlation of defects with the same sign shows a peak around $8\ \angstrom$ that characterizes the typical distance between the topological charges. This distance is compatible with the pattern in Fig.~\ref{fig:8}(a). We notice that this correlation is slightly stronger for defects with positive charge. 

By moving to higher frequency, we notice that the oscillations in the correlation between defects with same charge become weaker and the position of the first peak slowly moves to lower frequency (as for 2D amorphous solids, \cite{wu2023topology}). At high frequency, the correlation becomes more featureless, resembling more the structure of a gas, rather than a liquid of defects. Finally, the strong short-range correlation between negative and positive defects appears more robust with increasing frequency, even if its extent becomes more localized at short distances. At very large frequency, we still observe a weak correlation between positive and negative defects at short distance, followed by an almost flat decorrelated trend at intermediate and long distances. All the observed features in the positional correlation of the topological defects as a function of $\omega$ are extremely similar to those observed in glasses in \cite{wu2023topology,vaibhav2024experimental}. This suggests a possible universal character for these features that is not directly related to structural disorder.

Finally, in panel (d) of Fig.~\ref{fig:9} we study how their number $N_{\text{TD}}$ grows with $\omega$ for the case of a single edge dislocation in Fig.~\ref{fig:2} using two different grid sizes (black and red colors). Similar results have been obtained for the other cases. In the same figure, we also show the corresponding vibrational density of states (DOS) in blue color. We observe that the number of topological defects grows at low frequency following a power-law scaling $N_{\text{TD}}\propto \omega^2$, as found in athermal glasses \cite{wu2023topology}. Moreover, we find that the density of defects strongly correlates with the DOS at low frequency, as experimentally observed in a 2D colloidal glass in \cite{vaibhav2024experimental} (where, nevertheless, the scaling with frequency is linear because of finite temperature dissipative effects and the presence of unstable modes). Once again, we therefore suggest that these characteristics are not caused by disorder, as they appear both in glasses and crystals.

\section*{Outlook}

It has been recently claimed that dynamical defects defined using the topology of the vibrational modes \cite{wu2023topology} generalize the concept of structural defects in crystals to amorphous solids, providing a possible microscopic description of plasticity in disordered materials. Nevertheless, the connection between the two objects in defective crystalline structures has never been explored, leaving doubts about the aforementioned claim.

In this work, we have considered several types of structural defects in crystalline materials including dislocations, disclinations and Eshelby inclusions. In all these cases except for the hard Eshelby inclusion, we have found that the topology of the lowest vibrational mode well characterizes the presence and location of the defects in the structure of crystalline solids.

We have observed that dislocations are always associated to a negative net topological charge concentrated within their core. This supports the idea that anti-vortices, or more in general ``spots'' with negative net topological charge, have a privileged role as carriers of plasticity not only in glasses \cite{wu2023topology} but also in crystals. Disclinations can be also identified by a linear structure of topological dipoles that can be rationalized by visualizing the disclination as an array of dislocations.

Furthermore, soft Eshelby inclusions present a complicated pattern of topological defects inside their core, but the total topological charge associated to them is always equal to $q=-1$ as well. On the contrary, hard Eshelby inclusions are the only structural defects considered in this work that do not present any clear topological feature in the vibrational modes. Importantly, and consistent with our expectations, in the case of Eshelby inclusions (either soft or hard) we have found that the total topological charge of the whole system is always zero, confirming their non-topological nature. 

Moreover, we observed that this topological characterization performed withing the vibrational modes is particular efficient for the case of dislocations and isolated structural defects, while it becomes less clear in more complex scenarios. In particular, despite many features persist in the higher frequency vibrational modes, they are gradually hidden by the proliferation of many other topological defects (not related to structural defects and plasticity) that renders the identification of the structural defect and its location impossible using high frequency modes. By inspecting higher-frequency vibrational modes, we obtained that not only the number of defects grows with the frequency but their spatial structure becomes more and more disordered, resembling a gas-like environment. Finally, we have corroborated that the number of these topological defects nicely correlates with the vibrational density of state of the defective crystal considered. 

Our results suggest that many of the topological features of the vibrational modes recently observed in amorphous solids are actually universal, as they apply to defective crystals as well. Moreover, we conclude that there exists a strong connection between structural defects and topological dynamical defects in defective crystals. This poses on firmer ground the idea that these dynamical defects generalize the concepts of structural defects to amorphous materials, where the latter cannot be defined and identified anymore due to the underlying structural disorder.

To conclude, it is important to stress that not all topological defects relate to plasticity and, in our case, to topological structural defects in crystals. Indeed, many of them are of phononic nature or appear because of the complex vibrational pattern caused by the hybridization of plane waves and quasi-localised modes. It would be interesting to decouple these two contributions with the existing methods \cite{lerner2023boson,Moriel2024,mahajan2024revealing} and reveal which defects remain the relevant ones at higher frequency.

Apart from the spatial correlation between topological defects and plastic spots that have been confirmed in many instances, it is important to explore the dynamics of these topological defects in relation to the onset of plasticity and possible global instabilities such as yielding. This might directly connect to the idea of plasticity as ``screened elasticity'' \cite{kumar2024elasticity} and to the role of geometrical charges recently stressed in the literature \cite{PhysRevE.107.055004}. 

Last but not least, it is fundamental to generalize this analysis to 3D crystalline materials. In that direction, the definition of the winding number in Eq.~\eqref{wind} is not useful anymore and must be abandoned in favor of different topological invariants. One immediate option is to consider the Hedgehog topological charge, as recently proposed for 3D amorphous solids in \cite{bera2024hedgehog}.

We leave these interesting questions for the near future.

\section*{Acknowledgments} 

Useful discussions with A. Zaccone, A. Bera, Z. Wu, Amelia Y. Liu, W. Kob, M. Moshe, J. Zhang, Y. Wang, M. Pica Ciamarra, S. Mahajan, C. Jiang, Z. Zheng, T. Sirk and T. Petersen are gratefully acknowledged.
This work was financially supported by the Strategic Priority Research Program (grants nos. XDB0620103 and XDB0510301) and the Youth Innovation Promotion Association of Chinese Academy of Sciences, and the National Natural Science Foundation of China (grant no. 12472112). The numerical calculations in this study were carried out on the ORISE Super-computer. M.B. acknowledges the support of the Shanghai Municipal Science and Technology Major Project (grant no. 2019SHZDZX01) and the sponsorship from the Yangyang Development Fund.

%


\appendix

\section{Models of crystalline defects}
\label{methodsa}
\subsection*{Dislocations}

A BCC structure iron crystal model with x, y and z orientations of [1 1 1], [-1 1 0] and [-1 -1 2] was constructed for modeling full dislocation. For this purpose, an edge dislocation was constructed by superimposing two perfect crystals at the reduced coordinates (1/2, 1/2) of the simulation box. The slip plane of the edge dislocation is (-1 1 0), the dislocation line is along [-1 -1 2], and the direction of the Burgers vector is [1 1 1]. To get partial edge dislocations, a FCC structure aluminum crystal with x, y and z orientations of [1 1 0], [-1 1 1] and [1 -1 2] was constructed. Two edge dislocations with opposite directions of the Burgers vector were introduced at the reduced coordinates (1/4, 1/4) and (3/4, 3/4) of the simulation box along the y direction respectively. The slip plane of the edge dislocations is (-1 1 1), the dislocation line is along [1 -1 2], and the Burgers vector is $[1\ 1\ 0]a_{\mathrm{Al}}/2$, with $a_{\mathrm{Al}}= 4.046\ \angstrom$ the lattice constant of aluminum. After optimization aluminum crystal structure, one full edge dislocation of aluminum will decompose into two Shockley partial dislocations with Burgers vector $\langle1\ 1\ 2\rangle a_{\mathrm{Al}}/6$. Due to the high stacking energy of aluminum, the distance between two partials is short as such  the full dislocation is visible in the small simulation box of aluminum.

\subsection*{Disclination}

A BCC structure iron crystal model with x, y and z crystal orientations of [1 1 1], [-1 1 0] and [-1 -1 2] was constructed for modeling disclination. Here a wedge disclination is obtained by inserting an array of edge dislocations, which can be constructed by superimposing two crystals. Consequently, the disclination model can be regarded a grain boundary with very small tilt angle. The slip plane of a dislocation in the array is (-1 1 0), the dislocation line is along [-1 -1 2], and the direction of the Burgers vector is [1 1 1]. The distance $h$ between adjacent dislocations is approximately 16 $\angstrom$. The norm of the Frank vector of this wedge disclination is $\Omega=b/h=0.155\ \mathrm{rad}$, where $\vec{b}=[1\ 1\ 1]a_{\mathrm{Fe}}/2$ is the Burgers vector of the edge dislocation, $a_{\mathrm{Fe}}=2.855\ \angstrom$ is the lattice constant of iron. 

\subsection*{Eshelby inclusions}

For a hard inclusion, a BCC iron crystal model with x, y and z orientations of [1 1 1], [-1 1 0] and [-1 -1 2] was constructed firstly. Then, a carbon cylinder of radius 15 $\angstrom$ is inserted in the geometric center of the xy-plane along the z-axis. The carbon crystal is diamond lattice. On the contrary, for soft inclusion, a carbon model of diamond lattice was constructed as the matrix material. And a iron cylinder of radius 12 $\angstrom$ is placed in the geometric center of the xy-plane along the z-axis. The x, y and z orientations of the included iron are [1 1 1], [-1 1 0] and [-1 -1 2], respectively.

\section{Atomistic simulations}
\label{methodsb}
The empirical potential functions used for iron single crystal, aluminum single crystal and Eshelby inclusion are obtained from Refs.~\cite{doi:10.1080/14786430310001613264,PhysRevB.59.3393,PhysRevB.89.094102} respectively. For models of full edge dislocation and wedge disclination in iron, periodic boundary conditions are applied along the x and z directions, the y direction is fixed. For models of partial dislocation and inclusion, periodic boundary conditions are applied along all the x, y, and z directions. These boundary conditions were considered because of the models are quasi-2D in nature. The conjugate gradient method is used to relax atomic structures of the all defect models. The model of full edge dislocation in iron includes 10980 atoms. The model of partial dislocation in aluminum contains 14760 atoms. The model of wedge disclination in iron is of 12972 atoms. Finally, the model of hard inclusion has 8569 iron atoms and 765 carbon atoms, and the model of soft inclusion has 6907 carbon atoms and 133 iron atoms.

The software ATOMSK \cite{HIREL2015212} and LAMMPS \cite{lammps} were used for models construction and molecular dynamics simulation respectively. OVITO \cite{ovito} was used for visualization of the atomic configurations.

\section{Additional data}\label{add}
\begin{figure*}
    \centering
    \includegraphics[width=\linewidth]{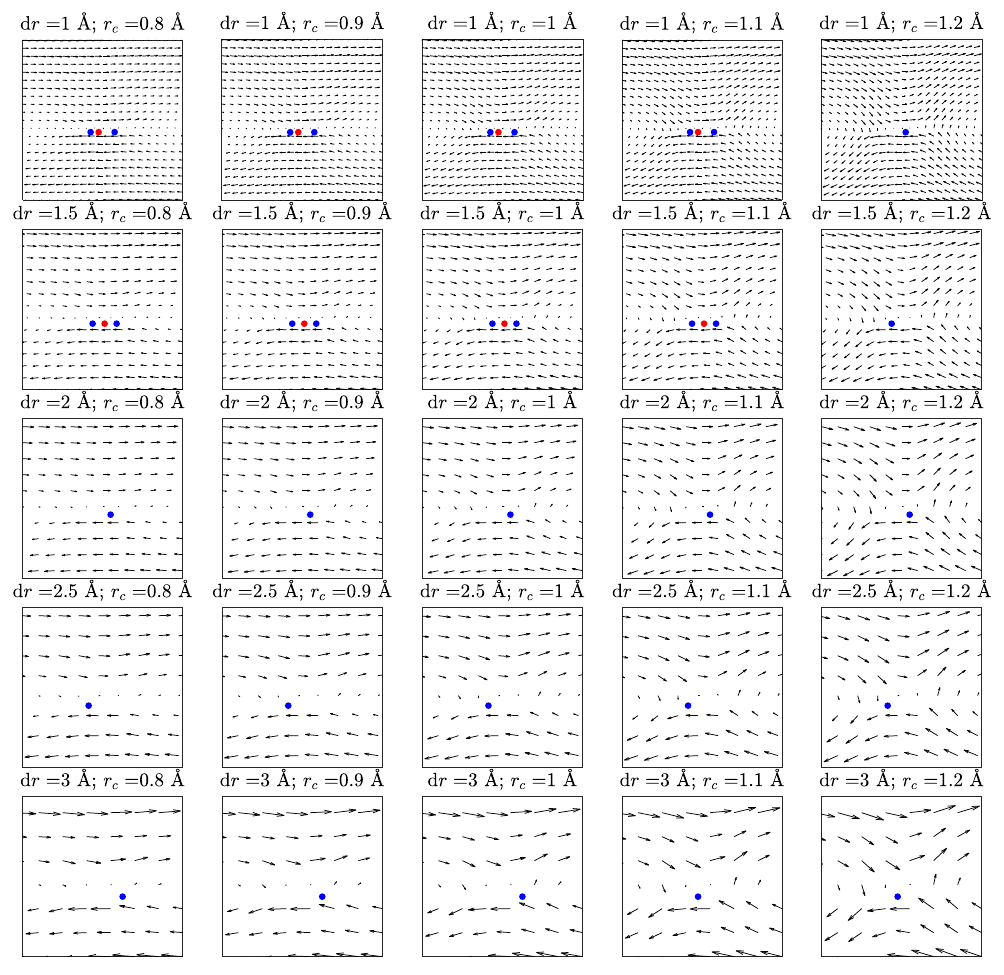}
    \caption{\textbf{Robustness of the topological features.} The smoothed-out eigenvector field and the topological defects calculated with different sizes of grid $\mathrm{d}r$ and cutoff distance $r_c$, in the case of the lowest frequency eigenvector ($\omega=0.63$ THz) for a single dislocation shown in Fig.~\ref{fig:2}. All quantities are shown in units of $\angstrom$.}
    \label{fig:10}
\end{figure*}

\begin{figure*}
    \centering
    \includegraphics[width=\linewidth]{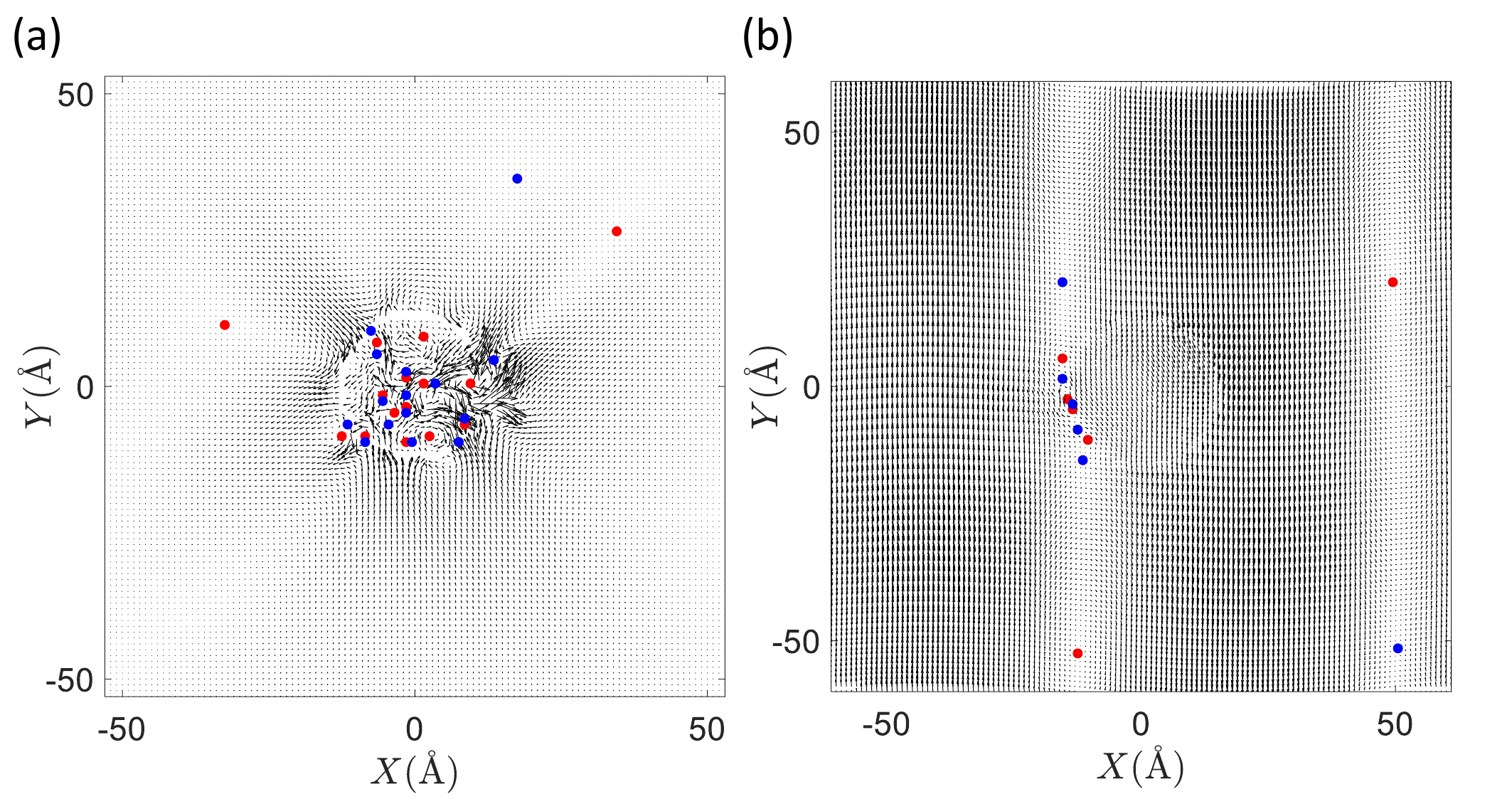}
    \caption{\textbf{Higher vibrational modes for soft and hard inclusions.} Eigenvector and topological charge in soft and hard Eshelby inclusion at the third lowest frequency modes, \textit{i.e.}, panel (a) corresponds to $\omega= 4.40$ THz and panel (b) to $1.45$ THz, respectively.}
    \label{fig:11}
\end{figure*}

Additional data are shown in Figs.~\ref{fig:10}-\ref{fig:11}. In Fig.~\ref{fig:10} we analyze the robustness of our results by changing the grid size and the cutoff distance $r_c$ for the case of a single dislocation presented in Fig.~\ref{fig:2} in the main text.

In Fig.~\ref{fig:11}, we show the topological characteristics for the third lowest frequency modes in the case of soft (panel (a)) and hard (panel (b)) inclusions. In Fig.~\ref{fig:11}(a), we notice that the number of topological dipoles inside the soft inclusions grows. The total topological charge of the soft inclusion is always $q=-1$ and the total system is always neutral. On the other hand, in Fig.~\ref{fig:11}(b) we notice that all the topological defects are trivial and of phononic nature, living on two parallel vertical bands located at $X\approx -10\ \angstrom$ and $X\approx 50\ \angstrom$. The band at $X\approx -10\ \angstrom$ is slightly distorted in the vicinity of the inclusion. The total system is still topological neutral.

\end{document}